\documentclass{article}
\usepackage{spconf,amsmath,graphicx,hyperref,booktabs,multirow, cite, amssymb}
\usepackage{orcidlink}
\usepackage{pgfplots}
\usepgfplotslibrary{groupplots}
\usetikzlibrary{calc}
\pgfplotsset{compat=1.17}

\title{MIDIBack: Harmony-Aware Singing Pitch Correction via Joint Vocal-Accompaniment Symbolic Modeling}
\name{Joaquim Cavalcante\orcidlink{0009-0002-4305-1900}$^{1}$ \enskip 
Yicheng Gu\orcidlink{0009-0001-7819-5667}$^{2}$ \enskip 
Adriel Trajano\orcidlink{0009-0003-8239-3302}$^{3}$ \enskip 
Yuri de Malheiros\orcidlink{0000-0002-7779-0288}$^{1}$ \enskip 
Thais Gaudencio\orcidlink{0000-0002-6608-4900}$^{1}$}
\address{$^{1}$ Centro de Informática, Federal University of Paraíba, João Pessoa, Brazil \\
      $^{2}$ Acoustic Lab, Aalto University, Espoo, Finland \\
      $^{3}$ MIR Research, Moises AI, João Pessoa, Brazil}
\begin{document}
%
\maketitle
\begin{abstract}
Automatic pitch correction (APC) requires distinguishing the unintended intonation errors from expressive pitch variation.  
Existing systems either lack explicit harmonic modeling, as vocal-only methods do, or do not directly use the note-level polyphonic context. 
Therefore, we propose MIDIBack, a note-level APC framework that jointly models the vocal and accompaniment events in a shared OctupleMIDI sequence. 
We evaluate MIDIBack under 6 note corruption regimes, including global outshift, learned note-dependent detuning, uniform perturbations, and their combinations. 
The resulting model achieves 78.6\% overall raw pitch accuracy (RPA), and 81.5\% under combined global outshift and learned detuning. 
Removing the accompaniment conditioning reduces RPA from 81.5\% to 35.8\% in outshift, showing the effectiveness of accompaniment context. 
Case studies on accompaniment modulation further illustrate that vocal note predictions will dynamically adapt to the shifts in the backing harmony.


\end{abstract}
\section{Introduction}
\label{sec:intro}

In modern music production, automatic pitch correction (APC) is an essential tool that corrects out-of-tune singing voices into in-tune singing segments. 
However, automatically inferring the recording's intended notes based on the musical context remains a challenging problem. 
Real-world singing recordings inevitably entangle intended musical notes with unintended intonation errors and stylistic pitch deviations. 
For instance, trained singers can reproduce target notes with deviations without being perceived as out of tune~\cite{Sundberg1995Replicability}, 
since vocal pitch perception is closely coupled with timbre, and listeners accept wider deviation in voice than in fixed-pitch instruments~\cite{Watts01012008}. 
This makes the intonation errors difficult to distinguish from intentional artistic deviations based on acoustic pitch alone. 
Therefore, an APC system cannot rely solely on naive nearest-semitone quantization, but needs to predict the intended vocal notes based on the musical cues.

Early pitch correction tools are based on DSP algorithms like TD-PSOLA~\cite{psola} and WORLD~\cite{world}, including commercial plugins such as Auto-Tune\footnote{\url{https://www.antarestech.com/}} and Melodyne\footnote{\url{https://shop.celemony.com/}}. 
However, they rely solely on static scale quantization with rigid rules based on music theory, which require manual adjustment, as the model may fail under rapid harmonic modulations and non-diatonic tension notes. As machine learning methods develop, pitch-controllable neural vocoders~\cite{sifigan, pcnsf, periodgrad} and codecs~\cite{neurodyne, periodcodec, pitchflower} have been proposed to enhance this regime, but the problem remains: they are not decision-makers and require manually adjusted vocal MIDI targets~\cite{karatuner, diffpitcher, stylepitcher}. 
Recent advancements in deep learning-based APC systems aim to solve this problem, yet remain limited in polyphonic and harmonic awareness. 
In particular, DDPC~\cite{DDPC} utilizes the acoustic CQT spectra as the condition, ignoring the rich note-level polyphonic context in the backing track. Meanwhile, BERT-APC~\cite{BertAPC2025} employs symbolic language models~\cite{MusicBERT} conditioned on vocal notes and coarse metrical and temporal metadata, leaving the rich harmonic cues in the accompaniment unexploited.

To bridge these gaps, this paper proposes MIDIBack, a note-level APC framework that jointly models the vocal and accompaniment events in a unified OctupleMIDI sequence, enabling the model to capture instantaneous polyphonic and chord progressions based on the backing track context. 
We conduct experiments under six pitch corruption regimes to illustrate its effectiveness, including global outshift, learned note-dependent detuning, uniform perturbations, and their combinations. 
Experimental results illustrate that MIDIBack achieves competitive results compared with all baselines, with 78.6\% raw pitch accuracy (RPA) in overall evaluation, $91.3\%$ RPA on clean transcriptions, and $81.5\%$ RPA under severe corruption settings.
Ablation studies are conducted on training regimes, accompaniment conditioning, and backbone architectures to confirm their effectiveness. 
We additionally conduct case studies on accompaniment modulation to show that the pitch correction decisions will be actively affected and adapted to the shifts in the backing harmony.







\begin{figure*}[t]
    \centering
    \includegraphics[width=0.95\textwidth]{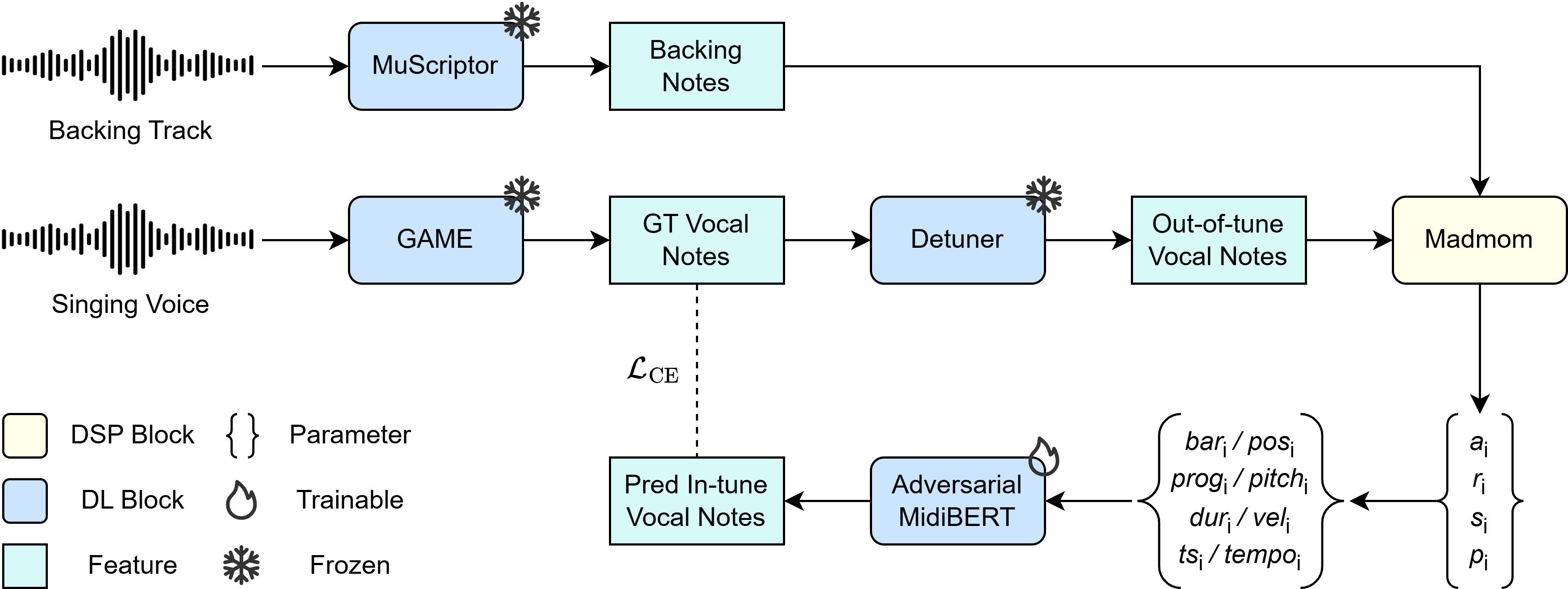}
    \caption{Overall architecture and training pipeline of the proposed MIDIBack system.}
    \label{fig:intuition}
\end{figure*}

\section{Methodology}
\label{sec:methodology}
\begin{table*}[t]
\centering
\caption{Objective evaluation results between MIDIBack and baseline models across six corruption regimes.
Raw Pitch Accuracy (RPA), Correction Rate (Corr), and Corruption Rate (Corp) are reported in percent (\%), and Mean Absolute Error (MAE) is measured in semitones (st).
Metrics are grouped as \textbf{RPA\,/\,MAE} ($\uparrow$\,/\,$\downarrow$) and \textbf{Corr\,/\,Corp} ($\uparrow$\,/\,$\downarrow$).
Best results within each training setup group (excluding Corrupted Input) are highlighted in \textbf{bold}.
Models marked under \emph{Uniform-Only} and \emph{Detune-Only} mean that they are trained on their native corruptions following the original papers of DDPC and BERT-APC, respectively.}
\label{tab:results}
\setlength\tabcolsep{1.5pt}
\resizebox{\linewidth}{!}{
\begin{tabular}{lcccccccccccc}
\toprule
\multirow{2}{*}{\textbf{System}} & \textbf{Overall} & \textbf{Clean} & \multicolumn{2}{c}{\textbf{Outshift}} & \multicolumn{2}{c}{\textbf{Detune}} & \multicolumn{2}{c}{\textbf{Uniform}} & \multicolumn{2}{c}{\textbf{Outshift + Detune}} & \multicolumn{2}{c}{\textbf{Outshift + Uniform}} \\
\cmidrule(lr){2-2} \cmidrule(lr){3-3} \cmidrule(lr){4-5} \cmidrule(lr){6-7} \cmidrule(lr){8-9} \cmidrule(lr){10-11} \cmidrule(lr){12-13}
& \textbf{RPA} & \textbf{RPA\,/\,Corp} & \textbf{RPA\,/\,MAE} & \textbf{Corr\,/\,Corp} & \textbf{RPA\,/\,MAE} & \textbf{Corr\,/\,Corp} & \textbf{RPA\,/\,MAE} & \textbf{Corr\,/\,Corp} & \textbf{RPA\,/\,MAE} & \textbf{Corr\,/\,Corp} & \textbf{RPA\,/\,MAE} & \textbf{Corr\,/\,Corp} \\
\midrule
Corrupted Input & $57.1$ & $100.0$ / $0.0$ & $32.2$ / $0.68$ & $0.0$ / $0.0$ & $76.6$ / $0.23$ & $0.0$ / $0.0$ & $50.0$ / $0.50$ & $0.0$ / $0.0$ & $43.6$ / $0.59$ & $0.0$ / $0.0$ & $40.5$ / $0.69$ & $0.0$ / $0.0$ \\
\midrule
\multicolumn{13}{l}{\emph{Uniform-Only}} \\
DDPC & $\mathbf{72.4}$ & $86.1$ / $13.9$ & $\mathbf{71.3}$ / $\mathbf{0.29}$ & $\mathbf{65.2}$ / $15.8$ & $\mathbf{80.8}$ / $\mathbf{0.19}$ & $\mathbf{64.8}$ / $14.3$ & $\mathbf{74.6}$ / $\mathbf{0.26}$ & $\mathbf{63.7}$ / $\mathbf{14.6}$ & $\mathbf{63.0}$ / $\mathbf{0.39}$ & $\mathbf{45.3}$ / $14.2$ & $\mathbf{58.5}$ / $\mathbf{0.46}$ & $\mathbf{40.4}$ / $\mathbf{15.0}$ \\
MIDIBack (Ours) & $69.8$ & $\mathbf{95.2}$ / $\mathbf{4.8}$ & $57.9$ / $0.45$ & $45.2$ / $\mathbf{15.2}$ & $\mathbf{80.8}$ / $0.20$ & $57.3$ / $\mathbf{12.0}$ & $73.1$ / $0.32$ & $62.7$ / $16.5$ & $57.5$ / $0.52$ & $31.8$ / $\mathbf{9.3}$ & $54.2$ / $0.66$ & $34.4$ / $16.7$ \\
\midrule
\multicolumn{13}{l}{\emph{Detune-Only}} \\
BERT-APC & $58.1$ & $\mathbf{98.6}$ / $\mathbf{1.4}$ & $33.7$ / $0.67$ & $\mathbf{25.6}$ / $49.2$ & $\mathbf{95.3}$ / $\mathbf{0.05}$ & $\mathbf{86.6}$ / $\mathbf{2.0}$ & $48.9$ / $0.52$ & $\mathbf{22.0}$ / $24.3$ & $33.9$ / $0.67$ & $\mathbf{9.7}$ / $34.8$ & $38.1$ / $0.74$ & $\mathbf{11.2}$ / $22.4$ \\
MIDIBack (Ours) & $\mathbf{59.3}$ & $97.1$ / $2.9$ & $\mathbf{36.2}$ / $\mathbf{0.65}$ & $18.8$ / $\mathbf{27.0}$ & $92.3$ / $0.08$ & $74.4$ / $2.2$ & $\mathbf{50.9}$ / $\mathbf{0.50}$ & $19.7$ / $\mathbf{18.0}$ & $\mathbf{39.4}$ / $\mathbf{0.62}$ & $7.7$ / $\mathbf{19.5}$ & $\mathbf{40.0}$ / $\mathbf{0.72}$ & $10.1$ / $\mathbf{16.2}$ \\
\bottomrule
\end{tabular}
}
\vspace{-14pt}
\end{table*}

\begin{table*}[t]
\centering
\caption{Ablation of training corruption combinations on MIDIBack.
Raw Pitch Accuracy (RPA), Correction Rate (Corr), and Corruption Rate (Corp) are reported in percent (\%), and Mean Absolute Error (MAE) is measured in semitones (st).
Metrics are grouped as \textbf{RPA\,/\,MAE} ($\uparrow$\,/\,$\downarrow$) and \textbf{Corr\,/\,Corp} ($\uparrow$\,/\,$\downarrow$).
Best results (excluding Corrupted Input) are highlighted in \textbf{bold}.
}
\label{tab:results_schedules}
\setlength\tabcolsep{1pt}
\resizebox{\linewidth}{!}{
\begin{tabular}{lcccccccccccc}
\toprule
\multirow{2}{*}{\textbf{System}} & \textbf{Overall} & \textbf{Clean} & \multicolumn{2}{c}{\textbf{Outshift}} & \multicolumn{2}{c}{\textbf{Detune}} & \multicolumn{2}{c}{\textbf{Uniform}} & \multicolumn{2}{c}{\textbf{Outshift + Detune}} & \multicolumn{2}{c}{\textbf{Outshift + Uniform}} \\
\cmidrule(lr){2-2} \cmidrule(lr){3-3} \cmidrule(lr){4-5} \cmidrule(lr){6-7} \cmidrule(lr){8-9} \cmidrule(lr){10-11} \cmidrule(lr){12-13}
& \textbf{RPA} & \textbf{RPA\,/\,Corp} & \textbf{RPA\,/\,MAE} & \textbf{Corr\,/\,Corp} & \textbf{RPA\,/\,MAE} & \textbf{Corr\,/\,Corp} & \textbf{RPA\,/\,MAE} & \textbf{Corr\,/\,Corp} & \textbf{RPA\,/\,MAE} & \textbf{Corr\,/\,Corp} & \textbf{RPA\,/\,MAE} & \textbf{Corr\,/\,Corp}  \\
\midrule
Corrupted Input & $57.1$ & $100.0$ / $0.0$ & $32.2$ / $0.68$ & $0.0$ / $0.0$ & $76.6$ / $0.23$ & $0.0$ / $0.0$ & $50.0$ / $0.50$ & $0.0$ / $0.0$ & $43.6$ / $0.59$ & $0.0$ / $0.0$ & $40.5$ / $0.69$ & $0.0$ / $0.0$ \\
\midrule
Uniform-Only & $68.7$ & $96.1$ / $3.9$ & $54.2$ / $0.48$ & $39.1$ / $\mathbf{13.9}$ & $80.7$ / $0.20$ & $54.4$ / $11.2$ & $\mathbf{73.1}$ / $\mathbf{0.32}$ & $64.1$ / $\mathbf{18.0}$ & $54.3$ / $0.54$ & $25.6$ / $\mathbf{8.7}$ & $53.8$ / $0.66$ & $34.9$ / $18.5$ \\
Detune-Only & $59.3$ & $\mathbf{97.1}$ / $\mathbf{2.9}$ & $36.2$ / $0.64$ & $18.8$ / $27.0$ & $\mathbf{92.3}$ / $\mathbf{0.08}$ & $74.4$ / $\mathbf{2.2}$ & $50.9$ / $0.50$ & $19.7$ / $\mathbf{18.0}$ & $39.4$ / $0.62$ & $7.7$ / $19.5$ & $40.0$ / $0.72$ & $10.1$ / $\mathbf{16.2}$ \\
Outshift-Only & $73.9$ & $91.3$ / $8.7$ & $\mathbf{85.0}$ / $\mathbf{0.18}$ & $\mathbf{86.1}$ / $17.1$ & $82.7$ / $0.18$ & $64.8$ / $11.8$ & $59.0$ / $0.48$ & $52.2$ / $34.1$ & $75.7$ / $0.34$ & $67.8$ / $14.1$ & $49.4$ / $0.68$ & $38.0$ / $33.7$ \\
Uniform + Detune & $69.6$ & $95.5$ / $4.5$ & $53.7$ / $0.49$ & $40.7$ / $18.9$ & $88.1$ / $0.12$ & $74.3$ / $7.6$ & $71.0$ / $0.35$ & $60.9$ / $18.8$ & $55.5$ / $0.49$ & $30.8$ / $12.4$ & $53.7$ / $0.66$ & $35.1$ / $19.0$ \\
Uniform + Outshift & $77.4$ & $92.3$ / $7.7$ & $82.1$ / $0.21$ & $82.0$ / $17.8$ & $81.5$ / $0.19$ & $67.0$ / $14.1$ & $71.4$ / $0.36$ & $\mathbf{65.2}$ / $22.3$ & $76.6$ / $0.31$ & $67.4$ / $11.6$ & $60.7$ / $0.57$ & $49.0$ / $22.2$ \\
Detune + Outshift & $75.0$ & $91.7$ / $8.3$ & $83.7$ / $0.21$ & $84.0$ / $17.0$ & $88.4$ / $0.12$ & $\mathbf{80.8}$ / $9.3$ & $55.6$ / $0.54$ & $46.4$ / $35.3$ & $\mathbf{82.8}$ / $\mathbf{0.20}$ & $\mathbf{80.0}$ / $13.7$ & $47.6$ / $0.71$ & $36.6$ / $36.4$ \\
Uniform + Detune + Outshift & $\mathbf{78.6}$ & $91.3$ / $8.7$ & $81.5$ / $0.22$ & $81.0$ / $17.7$ & $86.2$ / $0.14$ & $75.0$ / $10.3$ & $69.2$ / $0.40$ & $61.0$ / $22.6$ & $81.5$ / $0.22$ & $77.4$ / $13.2$ & $\mathbf{62.0}$ / $\mathbf{0.56}$ & $\mathbf{51.3}$ / $22.3$ \\
\bottomrule
\end{tabular}
}
\vspace{-16pt}
\end{table*}

\begin{table*}[t]
\centering
\caption{Architecture ablation on ``Uniform + Detune + Outshift'' regime. 
Raw Pitch Accuracy (RPA), Correction Rate (Corr), and Corruption Rate (Corp) are reported in percent (\%), and Mean Absolute Error (MAE) is measured in semitones (st).
Metrics are grouped as \textbf{RPA\,/\,MAE} ($\uparrow$\,/\,$\downarrow$) and \textbf{Corr\,/\,Corp} ($\uparrow$\,/\,$\downarrow$).
Best results (excluding Corrupted Input) are highlighted in \textbf{bold}.
}
\label{tab:results_architecture}
\setlength\tabcolsep{2pt}
\resizebox{\linewidth}{!}{
\begin{tabular}{lccccccccccc}
\toprule
\multirow{2}{*}{\textbf{System}} & \textbf{Clean} & \multicolumn{2}{c}{\textbf{Outshift}} & \multicolumn{2}{c}{\textbf{Detune}} & \multicolumn{2}{c}{\textbf{Uniform}} & \multicolumn{2}{c}{\textbf{Outshift + Detune}} & \multicolumn{2}{c}{\textbf{Outshift + Uniform}} \\
\cmidrule(lr){2-2} \cmidrule(lr){3-4} \cmidrule(lr){5-6} \cmidrule(lr){7-8} \cmidrule(lr){9-10} \cmidrule(lr){11-12}
& \textbf{RPA\,/\,Corp} & \textbf{RPA\,/\,MAE} & \textbf{Corr\,/\,Corp} & \textbf{RPA\,/\,MAE} & \textbf{Corr\,/\,Corp} & \textbf{RPA\,/\,MAE} & \textbf{Corr\,/\,Corp} & \textbf{RPA\,/\,MAE} & \textbf{Corr\,/\,Corp} & \textbf{RPA\,/\,MAE} & \textbf{Corr\,/\,Corp} \\
\midrule
Corrupted Input & $100.0$ / $0.0$ & $32.2$ / $0.68$ & $0.0$ / $0.0$ & $76.6$ / $0.23$ & $0.0$ / $0.0$ & $50.0$ / $0.50$ & $0.0$ / $0.0$ & $43.6$ / $0.59$ & $0.0$ / $0.0$ & $40.5$ / $0.69$ & $0.0$ / $0.0$ \\
\midrule
MIDIBack & $91.3$ / $8.7$ & $\mathbf{81.5}$ / $\mathbf{0.22}$ & $\mathbf{81.0}$ / $\mathbf{17.7}$ & $\mathbf{86.2}$ / $\mathbf{0.14}$ & $\mathbf{75.0}$ / $\mathbf{10.3}$ & $69.2$ / $0.40$ & $61.0$ / $22.6$ & $\mathbf{81.5}$ / $\mathbf{0.22}$ & $\mathbf{77.4}$ / $13.2$ & $\mathbf{62.0}$ / $\mathbf{0.56}$ & $\mathbf{51.3}$ / $\mathbf{22.3}$ \\
w/o accompaniment & $\mathbf{94.6}$ / $\mathbf{5.4}$ & $35.8$ / $0.66$ & $24.2$ / $39.8$ & $84.5$ / $0.16$ & $67.9$ / $10.5$ & $53.0$ / $0.50$ & $36.6$ / $30.6$ & $38.0$ / $0.65$ & $13.9$ / $30.8$ & $37.7$ / $0.74$ & $17.4$ / $32.4$ \\
w/o advMidiBERT & $92.4$ / $7.6$ & $75.8$ / $0.32$ & $73.1$ / $18.4$ & $84.9$ / $0.15$ & $74.8$ / $12.0$ & $\mathbf{72.3}$ / $\mathbf{0.35}$ & $\mathbf{67.0}$ / $\mathbf{22.3}$ & $78.3$ / $0.29$ & $71.0$ / $\mathbf{12.2}$ & $58.2$ / $0.64$ & $45.7$ / $23.5$ \\
\bottomrule
\end{tabular}
}
\vspace{-16pt}
\end{table*}

MIDIBack treats APC as context-aware vocal note estimation over a unified multi-track OctupleMIDI sequence, as shown in Fig.~\ref{fig:intuition}. 
The system operates via two pre-trained symbolic transcription models, a detuner, and a trainable backbone.

\subsection{Multi-Track Symbolic Transcription}
\label{sec:method_transcription}
Given a singing recording and backing pair, the recording is segmented using~\footnote{\url{https://github.com/openvpi/audio-slicer}} and then transcribed into continuous vocal notes using GAME\footnote{\url{https://github.com/openvpi/GAME}}, yielding stationary pitch centers $\tilde{p}_i \in \mathbb{R}$. 
The backing audio is transcribed into symbolic notes via MuScriptor~\cite{muscriptor}. 
Both streams are merged and processed by Madmom~\cite{madmom} to extract the unified note-level temporal attributes: onset $a_i$, duration $r_i$, signature $s_i$, and pitch $p_i$.

\subsection{OctupleMIDI Representation}
\label{sec:method_representation}
The extracted singing and backing notes are then serialized into a single chronological OctupleMIDI sequence to comply with the backbone model scheme. 
In particular, each musical event is parameterized as an 8-tuple, as illustrated below:
\begin{equation}
\label{eq:octuple}
e_i = (\mathrm{bar}_i, \mathrm{pos}_i, \mathrm{prog}_i, \mathrm{pitch}_i, \mathrm{dur}_i, \mathrm{vel}_i, \mathrm{ts}_i, \mathrm{tempo}_i),
\end{equation}
which represents the bar index, intra-bar quantized position, instrument program, pitch, quantized duration, velocity, time signature, and tempo. In particular, $\mathrm{prog}_i=0$ denotes the vocal, and non-zero programs denote the backing tracks.

For continuous vocal pitches $\tilde{p}_i$, we interpolate adjacent semitone embeddings via $\alpha_i = \tilde{p}_i - \lfloor \tilde{p}_i \rfloor$ to obtain:
\begin{equation}
\label{eq:interp}
E_{\mathrm{pitch}}(\tilde{p}_i) = (1 - \alpha_i) E(\lfloor \tilde{p}_i \rfloor) + \alpha_i E(\lfloor \tilde{p}_i \rfloor + 1),
\end{equation}
which makes it possible to retain the microtonal intonation nuance without vocabulary expansion.

\subsection{Intonation Corruption}
\label{sec:method_corruption}
We apply specific rules to construct the corrupted out-of-tune vocal notes, including a frozen GRU detuner~\cite{BertAPC2025}, a global outshift, and uniform noise, which can be shown as follows:
\begin{equation}
\label{eq:corruption}
\bar{p}_i = \tilde{p}_i + b + \hat{\varepsilon}_i + u_i,
\end{equation}
where $b \sim \mathcal{U}(\pm[0.25, 1.0])$\,st represents a phrase-level key outshift, $\hat{\varepsilon}_i$ is an autoregressive note-to-note drift from the pre-trained detuner, and $u_i \sim \mathcal{U}(-1.0, 1.0)$\,st is a noise that provides uniform perturbation. Accompaniment notes remain uncorrupted, serving as the invariant musical anchor.

\subsection{Context-Aware Pitch Prediction}
\label{sec:method_cnpp}
We employ the Adversarial-MidiBERT~\cite{AdversarialMidiBERT2025} model as the trainable backbone, as its bidirectional architecture allows the model to see the in-tune context both before and after the out-of-tune area. 
Due to the scarcity of ground-truth score annotations in large-scale accompanied datasets, we adopt the rounded transcription centers $q_i = \mathrm{round}(\tilde{p}_i)$ (Sec.~\ref{sec:method_transcription}) as the pseudo ground-truth target. 
Therefore, the network is trained to invert the synthetic corruptions introduced in Eq.~\eqref{eq:corruption}.
During training, it is optimized via the cross-entropy loss against these targets strictly at vocal note positions:
\begin{equation}
\label{eq:ce}
\mathcal{L}_{\mathrm{CE}}
= -\frac{1}{N_v}\sum_{i=1}^{N_v}
\log P_\theta\!\bigl(q_i \mid \mathbf{e}_{\le M}\bigr),
\end{equation}
where $N_v$ represents the number of vocal notes and $M$ is the total sequence length. It is worth noting that the loss gradients from the backing token positions are blocked to ensure the accompaniment acts purely as a conditioning context.

\section{Experiments}
\vspace{-10pt}

We apply comprehensive evaluation, ablation studies, and case studies to show the effectiveness of the proposed system. Representative samples can be found on our demo page\footnote{\url{https://midiback.joaquimbreno.com/}}.

\vspace{-10pt}
\subsection{Experiment Setup}

\quad \hskip0.6em\relax \textbf{Datasets}: 
We trained MIDIBack using both vocal-only and accompanied singing voice datasets. 
For detuner, we use a data mixture comprising 49,730 clips (528.5 hours) from 9 datasets, including 
CSD~\cite{csd}, KiSing~\cite{kising}, Opencpop~\cite{Opencpop}, M4Singer~\cite{M4Singer}, MRSAudio~\cite{mrsaudio}, jaCappella~\cite{jacappella}, Korean Multisinger\footnote{\href{https://www.aihub.or.kr/aihubdata/data/view.do?dataSetSn=473}{Korean Multisinger}}, Korean Multitimbre\footnote{\href{https://www.aihub.or.kr/aihubdata/data/view.do?dataSetSn=465}{Korean Multitimbre}}, and SingStyle111~\cite{SingStyle}. 
For vocal pitch correction training, we use 2,356 multi-track songs (66.2 vocal hours) with both vocal and backing track audio, including MIR-1K~\cite{mir1k}, Sonovox\footnote{\url{https://sonovox.ai/}}, MoisesDB~\cite{moisesdb}, DSD100~\cite{DSD100}, and ccMixter~\cite{ccmixer}. 
Datasets are partitioned into a split of 80\%/10\%/10\% with a fixed random seed of 42 for training, validation, and testing, respectively. 
All audio recordings are resampled to 22.05\,kHz mono WAV files.

\textbf{Configurations}: 
MIDIBack is initialized on Adversarial-MidiBERT~\cite{AdversarialMidiBERT2025}, comprising a 12-layer Transformer encoder with a hidden dimension of $d=768$, 12 attention heads, and a maximum Octuple sequence length of 1024. The baselines are adapted from MusicBERT-base~\cite{MusicBERT} with the same config.

\textbf{Baselines}:
We compare MIDIBack against DDPC~\cite{DDPC}, a reference-based model that conditions on raw CQT spectra, and BERT-APC~\cite{BertAPC2025}, a reference-free system that predicts in-tune target notes based on the vocal context and metrical cues. Note that DDPC operates on the raw pitch, and we quantize its output into MIDI notes by rounding the pitch center.

\textbf{Training}:
The detuner is trained following the same scheme as in BERT-APC~\cite{BertAPC2025}, with 10\% highly detuned samples filtered to ensure stability. 
The vocal note correction model is trained using the AdamW optimizer with a base learning rate of $5\times10^{-6}$ for the encoder and $1\times10^{-3}$ for the prediction head, $(\beta_1, \beta_2) = (0.9, 0.98)$, weight decay of $0.01$, and a batch size of 24. A cosine annealing learning rate scheduler with $T_{\max}=500{,}000$ is utilized. 
Corruptions are applied progressively during training, in which epochs 0--19 run on clean input only, and after which the corruption probabilities anneal linearly from $0$ to $0.4$ in $11{,}000$ steps. 
All models are trained on an NVIDIA H100 using Hugging Face Accelerate in BF16 precision. All baselines are reproduced with the same datasets to ensure a fair comparison.

\textbf{Evaluation}:
We conduct evaluation under multiple note-corruption regimes and report Raw Pitch Accuracy (RPA, the percentage of predicted MIDI notes matching the rounded transcription MIDI targets, computed over all notes in the eval set), Mean Absolute Error (MAE) in semitones, Correction Rate (Corr, the ratio of notes originally wrong but fixed in the prediction), and Corruption Rate (Corp, the ratio of notes originally correct but got altered in the prediction).


\subsection{Experimental Results}
\label{sec:exp_results}

\vspace{-4pt}

We compare MIDIBack with baseline APC systems across six pitch-corruption scenarios under the same training regimes, as illustrated in Table~\ref{tab:results}. 
Compared with DDPC, MIDIBack achieves competitive accuracy on the Uniform ($73.1\%$ vs. $74.6\%$) and Detune ($80.8\%$) conditions, while performing better on clean inputs ($95.2\%$ vs. $86.1\%$, reducing Corp from $13.9\%$ to $4.8\%$). Compared with BERT-APC, MIDIBack has comparable results on the learned detuning ($92.3\%$) and clean ($97.1\%$) conditions while performing better on the Outshift ($36.2\%$) and compound conditions ($39.4\%$ and $40.0\%$).

\vspace{-10pt}

\begin{figure}[t]
\centering
\includegraphics[width=\columnwidth]{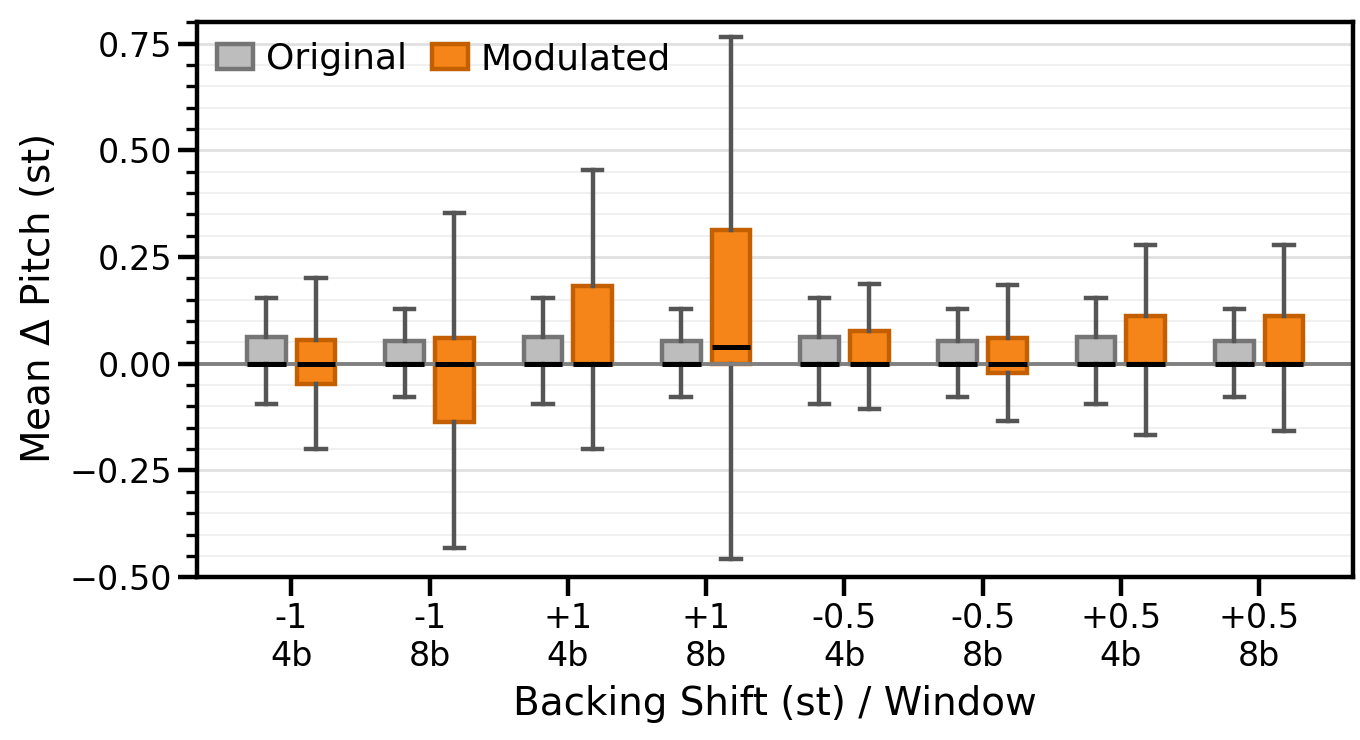}
\vspace{-24pt}
\caption{Boxplots of vocal pitch shift ($\Delta = \text{Pred} - \text{Input}$) when the input remains fixed at the accompaniment transcriptions are modulated by $\pm 0.5$ and $\pm 1.0$\,st. ``4b'' and ``8b'' mean the duration of modulation is 4- and 8-bar windows, respectively.}
\label{fig:vocal_change}
\vspace{-18pt}
\end{figure}

\subsection{Ablation Study}
\label{sec:ablation}

\vspace{-4pt}

We investigate the impact of different and compound training regimes on model performance across six pitch-corruption scenarios, as shown in Table~\ref{tab:results_schedules}. Specifically, models trained on single corruption scenarios tend to overfit their training regime. Pairwise combinations improve stability, with Detune + Outshift leading with the best result in compound evaluation ($82.8\%$ RPA, $80.0\%$ Corr). 
Integrating all scenarios yields the most balanced result, achieving the highest results on overall and Outshift + Uniform scenarios of $78.6\%$ and $62.0\%$.
\begin{figure}[tb]
\centering
\includegraphics[width=\columnwidth]{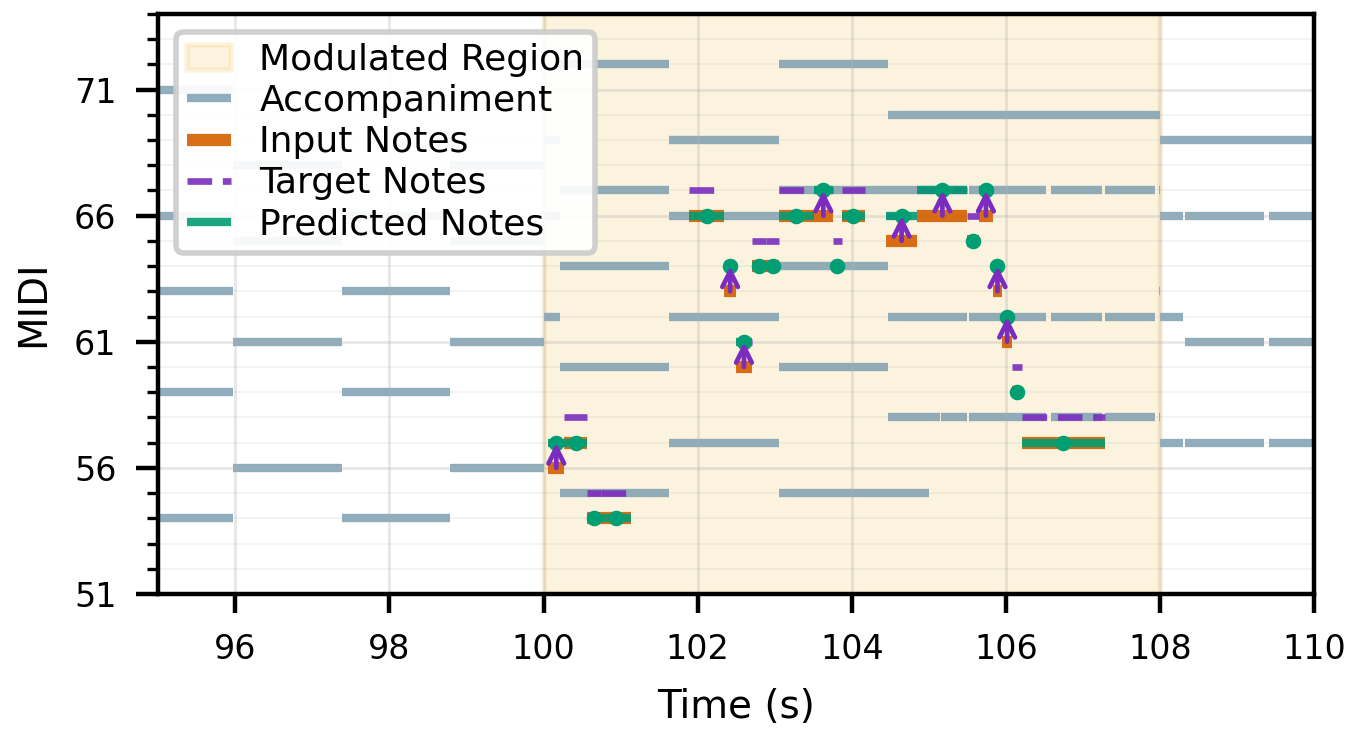}
\vspace{-24pt}
\caption{Vocal pitch correction in a region where the backing notes are modulated by $+1.0$\,st with unaltered vocal inputs.}
\label{fig:modulation_example}
\vspace{-18pt}
\end{figure}

We also validate the contribution of accompaniment note conditioning and the improved backbone architecture, as in Table~\ref{tab:results_architecture}. Removing the accompaniment hurts performance across all scenarios, especially the global-outshift condition, as the model will fail to capture the correct key. Substituting with MusicBERT leads to degradation in almost all settings, confirming the effectiveness of AdversarialMidiBERT. 

\subsection{Case Study}
\label{sec:case_study}
\vspace{-4pt}

We conduct controlled case studies to verify whether MIDIBack utilizes the symbolic harmonic and polyphonic context rather than solely memorizing melodic priors. 
We fix the vocal notes to the ground truth and modulate in-region backing notes by $\pm 0.5$ and $\pm 1.0$\,st over continuous 4- and 8-bar windows. 
As shown in Fig.~\ref{fig:vocal_change}, the predicted pitch residuals change along the direction of the modulation, with an amplified effect over longer context windows ($8$ bars). Figure~\ref{fig:modulation_example} additionally highlights a representative $+1.0$\,st phrase example, where the accompaniment shift actively guides vocal predictions toward the harmonically modulated targets, confirming that MIDIBack dynamically reasons over the symbolic context.

\vspace{-10pt}

\section{Conclusion}
\vspace{-6pt}

We introduced MIDIBack, a note-level APC framework that conditions vocal pitch prediction on symabolic backing tracks through a shared OctupleMIDI sequence.
Across 6 controlled corruption regimes, the complete model achieved the highest overall RPA among the evaluated training regimes. 
Removing accompaniment conditioning led to performance degradation, especially under global pitch offsets,
and controlled backing track modulation showed that vocal predictions dynamically adapt to the underlying harmonic changes.
All these findings indicate that the model reasons over harmonic context rather than relying exclusively on vocal pitch priors. 
Note that our evaluation is based on transcription-derived MIDI targets and simulated pitch corruptions, it mainly benchmarks corruption recovery rather than perceptual vocal tuning quality. 
Future work will evaluate under professionally annotated datasets and conduct subjective tests on end-to-end audio resynthesis.




\section{Acknowledgments}
\vspace{-6pt}
This study was financed in part by the Coordenação de Aperfeiçoamento de Pessoal de Nível Superior - Brasil (CAPES) - Finance Code 001, and in part by Moises AI.

\vspace{-8pt}

\bibliographystyle{IEEEbib}

\bibliography{refs}
\vspace{-8pt}

\end{document}